\documentclass{Interspeech}

\interspeechcameraready

\usepackage{multirow}
\usepackage{colortbl}
\usepackage[table,xcdraw]{xcolor}
\usepackage{tabularx}
\usepackage{todonotes}
\usepackage{graphicx}
\usepackage{cite}
\usepackage{float} 
\usepackage{hyperref}

\newcommand{\dbname}{AI4T\xspace}

\newcommand{\etal}{\textit{et al.}\xspace}

\title{Unmasking real-world audio deepfakes: A data-centric approach}

\author[affiliation={1}]{David}{Combei}
\author[affiliation={1,2}]{Adriana}{Stan}
\author[affiliation={2}]{Dan}{Oneata}
\author[affiliation={3}]{Nicolas}{Müller}
\author[affiliation={2}]{Horia}{Cucu}

\affiliation{}{Technical University of Cluj-Napoca}{Romania}
\affiliation{}{University ``Politehnica'' Bucharest}{Romania}
\affiliation{}{Fraunhofer AISEC}{Germany}
\email{david.combei@cs.utcluj.ro, adriana.stan@com.utcluj.ro, dan.oneata@gmail.com, nicolas.mueller@aisec.fraunhofer.de, horia.cucu@upb.ro}

\keywords{audio deepfake detection, data-centric, real-world deepfakes, antispoofing, data pruning, SSL}

\newcommand{\dan}[1]{\textcolor{black}{#1}}
\usepackage{comment}

\begin{document}
\maketitle

\begin{abstract}

   

The growing prevalence of real-world deepfakes presents a critical challenge for existing detection systems, which are often evaluated on datasets collected just for scientific purposes. To address this gap, we introduce a novel dataset of real-world audio deepfakes. Our analysis reveals that these real-world examples pose significant challenges, even for the most performant detection models. 
Rather than increasing model complexity or exhaustively search for a better alternative, in this work we focus on a data-centric paradigm, employing strategies like dataset curation, pruning, and augmentation to improve model robustness and generalization.

Through these methods, we achieve a 55\% relative reduction in EER on the In-the-Wild dataset, reaching an absolute EER of 1.7\%, and a 63\% reduction on our newly proposed real-world deepfakes dataset, \dbname. These results highlight the transformative potential of data-centric approaches in enhancing deepfake detection for real-world applications. Code and data available at: \url{https://github.com/davidcombei/AI4T}.
 
\end{abstract}

\vspace{-.2cm}
\section{Introduction}

\textit{``As you know our FTX exchange is going bankrupt, but
you should not panic[...] we have prepared a giveaway.
Just go to the site \dots'' 
}
We are assured by Sam Bankman-Fried (FTX CEO) in a viral clip posted on Twitter in November 2022.
Or are we?
The video turned out to be fake, but many were fooled by it and, consequently, lost money.
Such automatically generated clips (known as deepfakes) are permeating our society resulting in misinformation, scams and making us doubt the veracity of what we are seeing daily on the internet.

Deepfake detection aims to sidestep these risks by automatically identifying synthetically generated contented---video, images, audio.
As with any automatic approach, the essential component is data.
Data allows training deepfake detection methods, as well as measuring their progress.
As such, many datasets are continuously proposed \cite{asv19,for,asv21, timittts,odss,muller2024mlaad,muller2022does}.
These datasets follow the emergence of recent TTS systems~\cite{lyth2024natural,Gong2023ZMM-TTS,kharitonov2023speak}, or
aim to extend the number of speakers or languages \cite{odss,muller2024mlaad}.
However, these scientific (in the lab) datasets differ in an essential way from real-world deepfakes, i.e. the samples we encounter in the online/social media environment.
\textbf{Real-world deepfakes} are in most cases released in the online space to mislead.
Their development involves a human-in-the-loop approach:
the human examines the end result of the speech generator and adjusts its process to obtain an enhanced end-result.
The developer would listen to the generated sample and, if needed, update the parameters, input text, or change the generative system altogether--in a wider sense, they would curate one sample at a time.
This is in contrast with the \textbf{scientific deepfakes} where samples are produced in bulk
with minimal additional care towards quality and consistency. 

\begin{figure}
    \centering
    \setlength{\tabcolsep}{2pt}
    \begin{tabular}{ccccc}

        \href{https://www.youtube.com/watch?v=yz8go3S7CaI}{\raisebox{0pt}{
        \includegraphics[width=0.18\columnwidth]{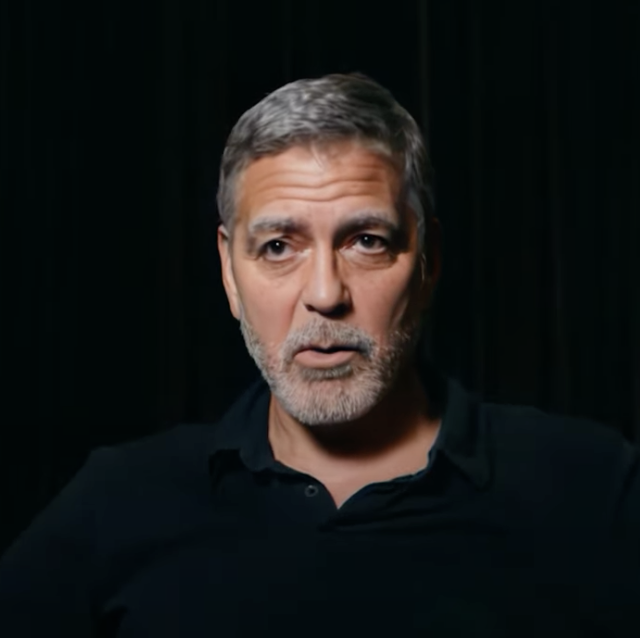}}} &

        \href{https://www.youtube.com/watch?v=IZg4YL2yaM0}{\raisebox{0pt}{
        \includegraphics[width=0.18\columnwidth]{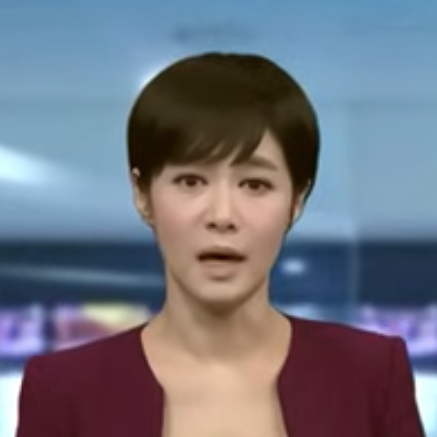}}} &

        \href{https://www.youtube.com/watch?v=c9tNqF6IQwY}{\raisebox{0pt}{
        \includegraphics[width=0.18\columnwidth]{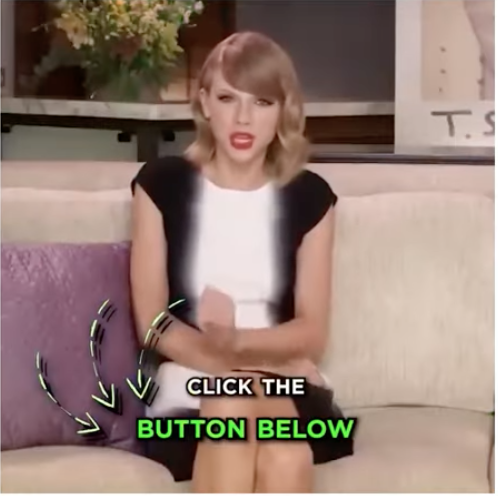}}} &

        \href{https://www.youtube.com/watch?v=GrkzMDYv5dY}{\raisebox{0pt}{
        \includegraphics[width=0.18\columnwidth]{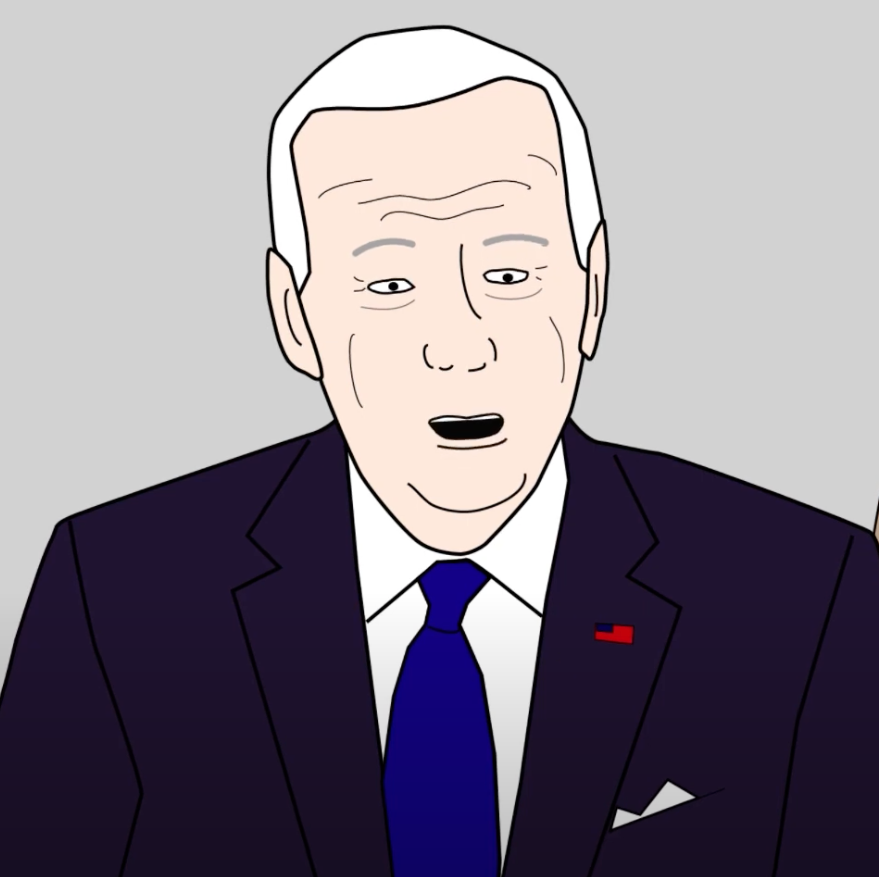}}} &

        \href{https://www.youtube.com/watch?v=X17yrEV5sl4}{\raisebox{0pt}{
        \includegraphics[width=0.18\columnwidth]{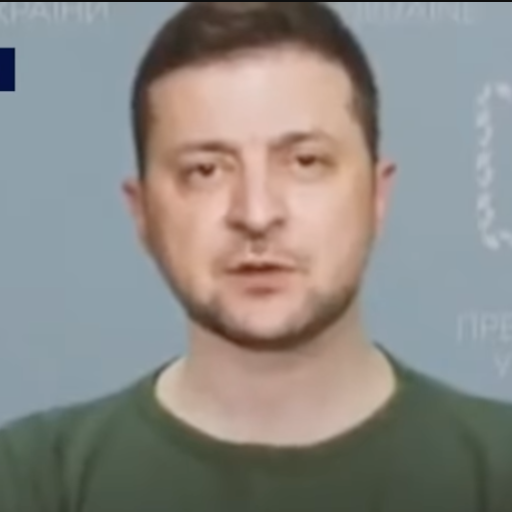}}} \\
        
        \includegraphics[width=0.18\columnwidth,height=0.5cm]{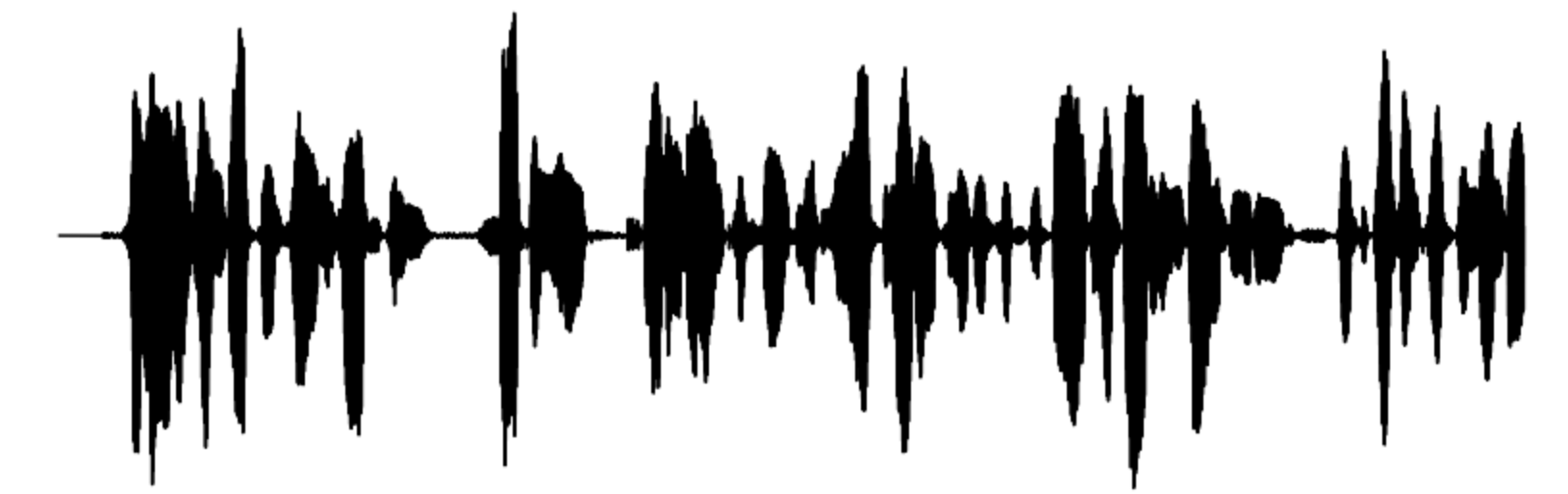} &
        \includegraphics[width=0.18\columnwidth,height=0.5cm]{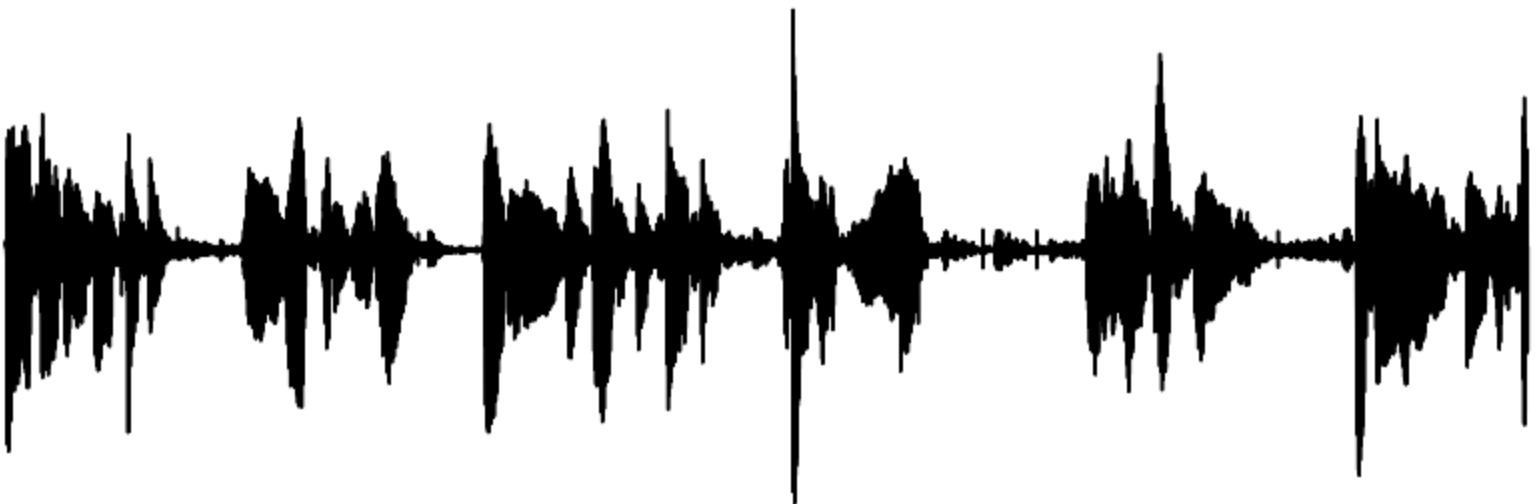} &
        \includegraphics[width=0.18\columnwidth,height=0.5cm]{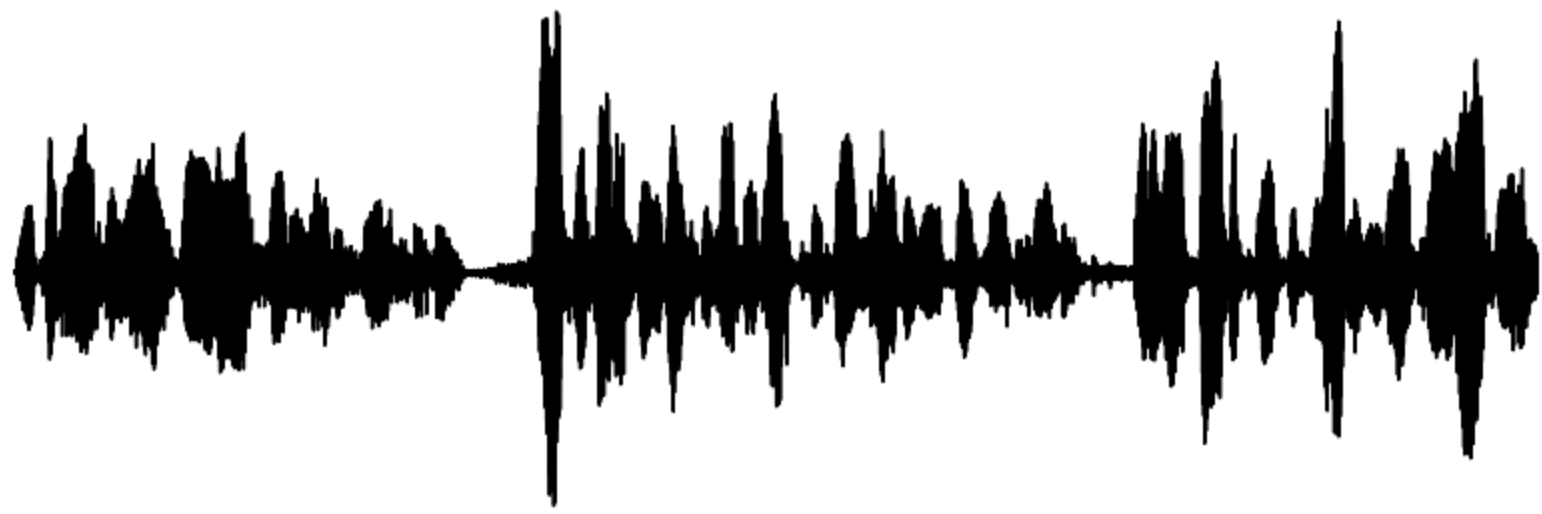} &
        \includegraphics[width=0.18\columnwidth,height=0.5cm]{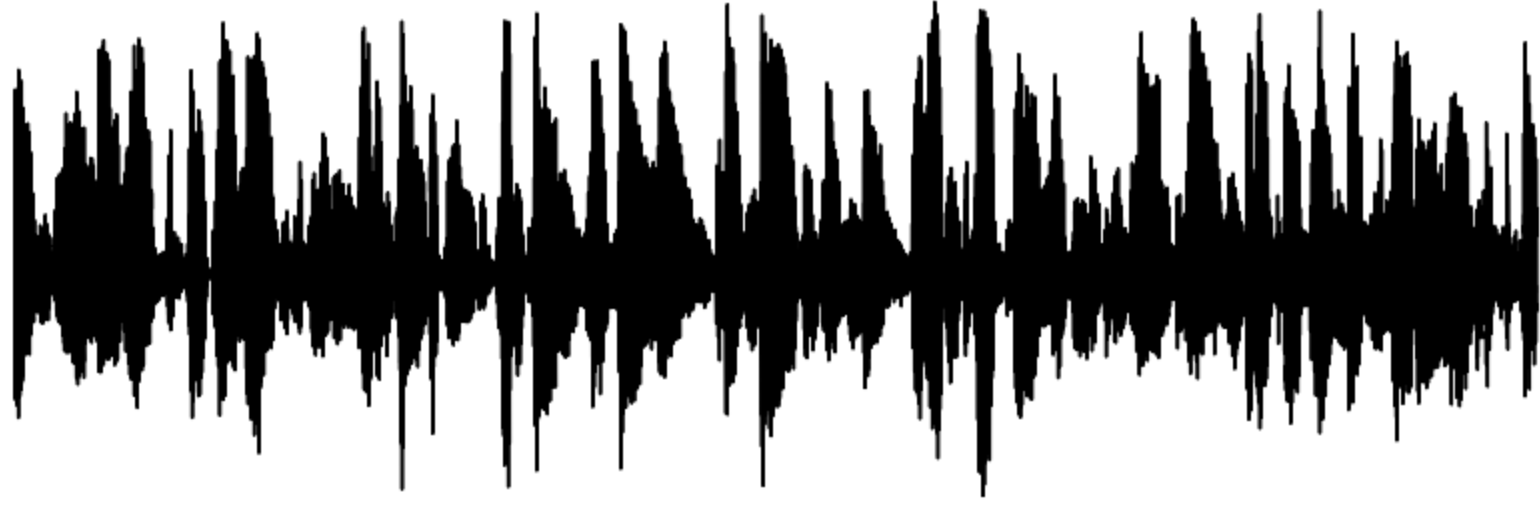} &
        \includegraphics[width=0.18\columnwidth,height=0.5cm]{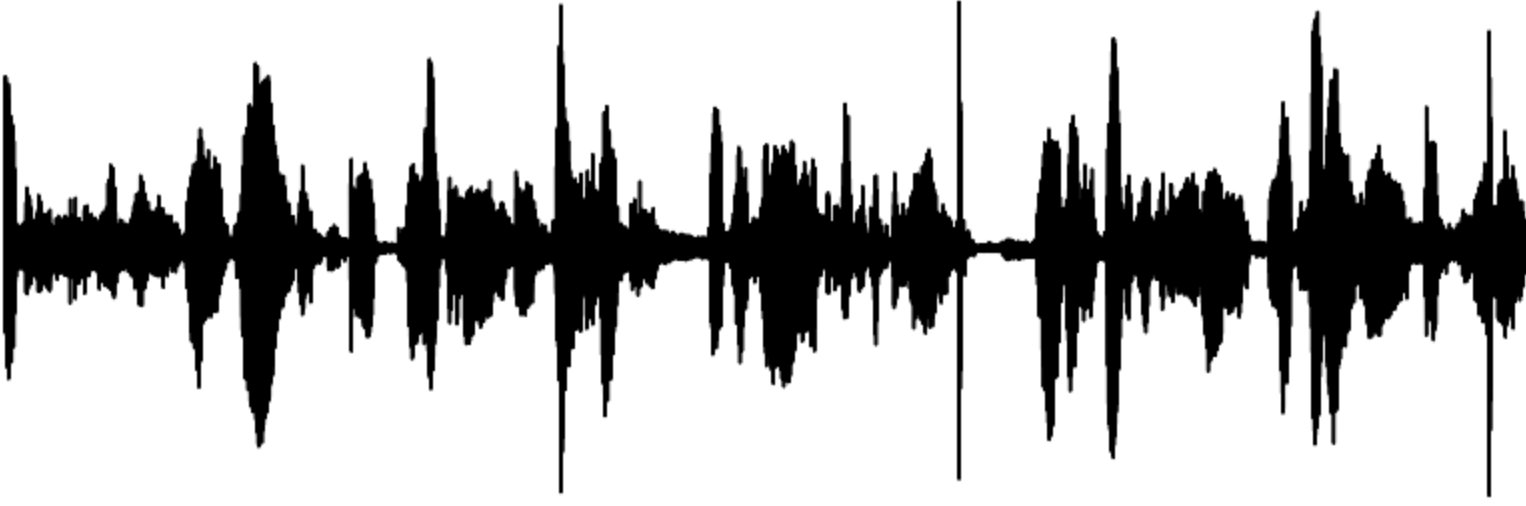} \\

        \href{https://www.youtube.com/watch?v=VLo_xIAiKzU}{\raisebox{0pt}{\includegraphics[width=0.18\columnwidth]{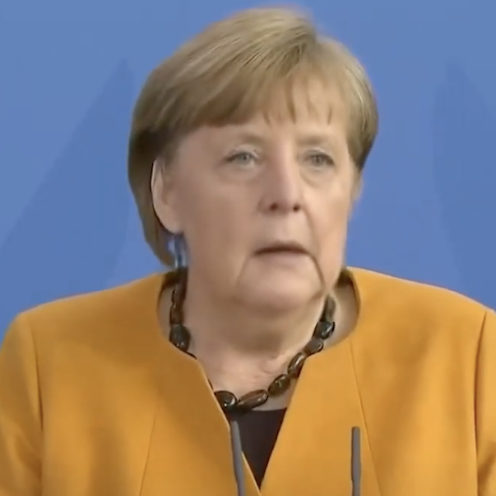}}} &
        
        \href{https://www.youtube.com/watch?v=e9hDan8vLTU}{\raisebox{0pt}{
        \includegraphics[width=0.18\columnwidth]{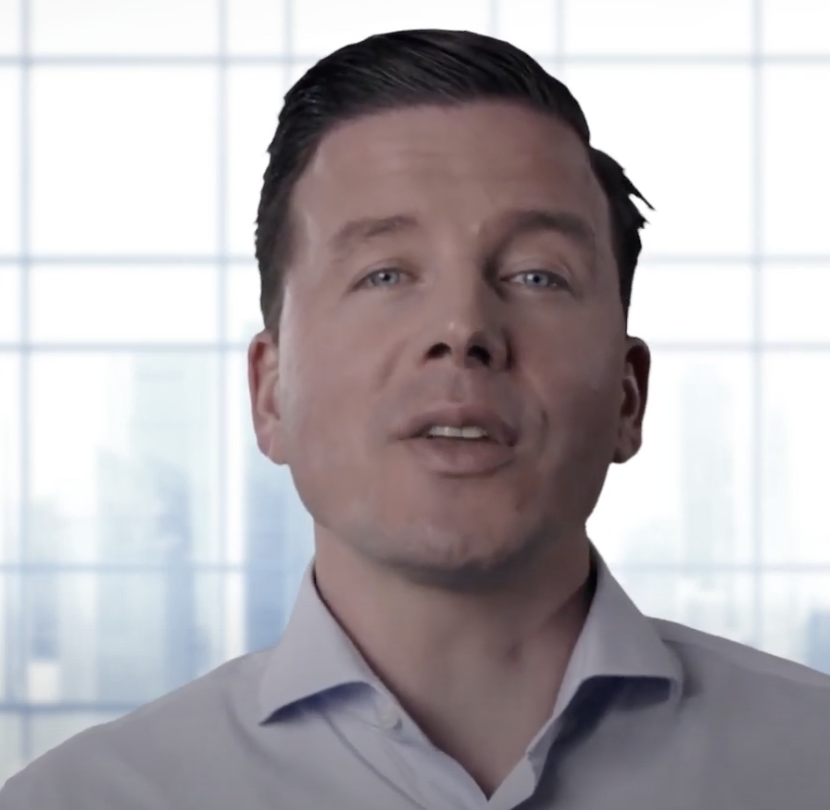}}} &

        \href{https://www.youtube.com/watch?v=cQ54GDm1eL0}{\raisebox{0pt}{
        \includegraphics[width=0.18\columnwidth]{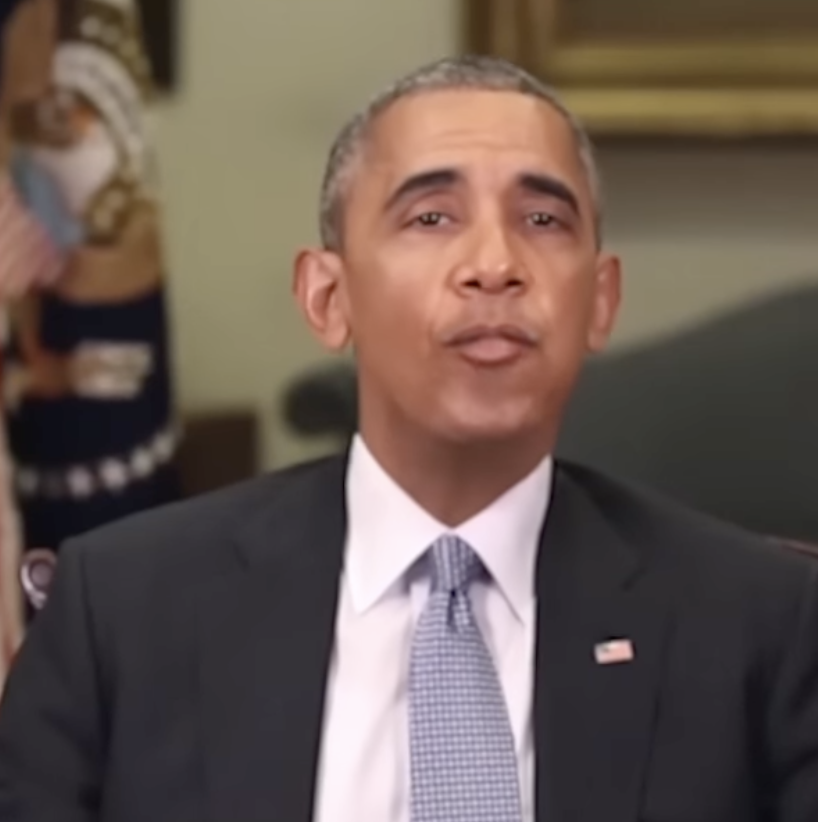}}} &

        \href{https://www.youtube.com/watch?v=hWSHh0kJz3s}{\raisebox{0pt}{
        \includegraphics[width=0.18\columnwidth]{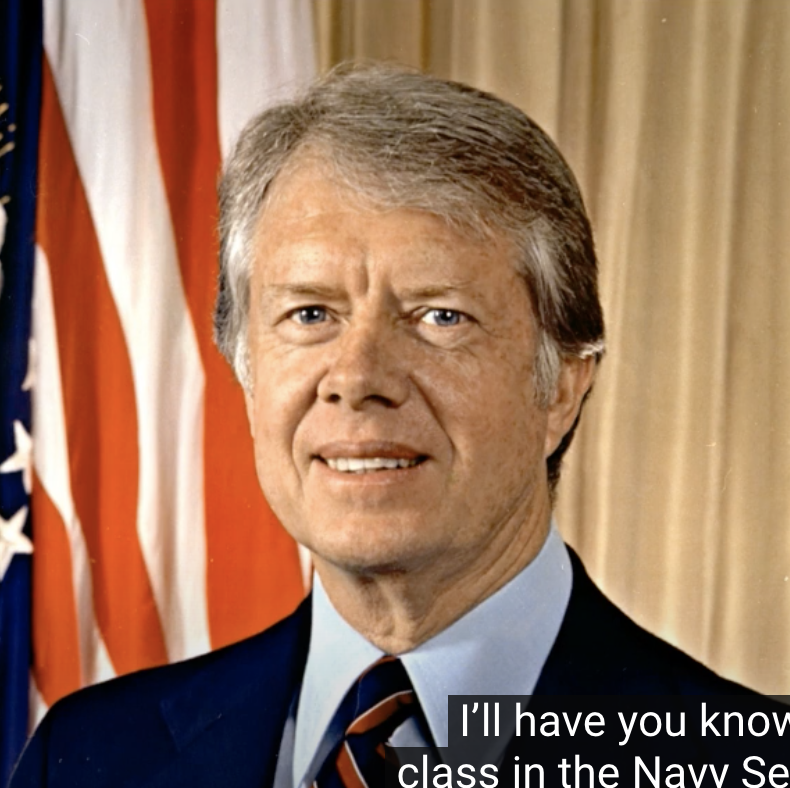}}} &

        \href{https://www.youtube.com/watch?v=T5FYqO8d6io}{\raisebox{0pt}{
        \includegraphics[width=0.18\columnwidth]{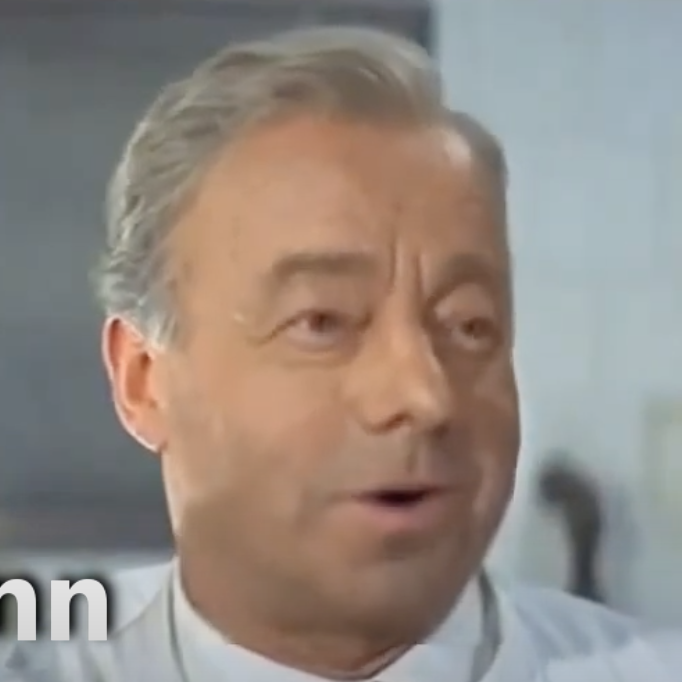}}} \\
        \includegraphics[width=0.18\columnwidth,height=0.5cm]{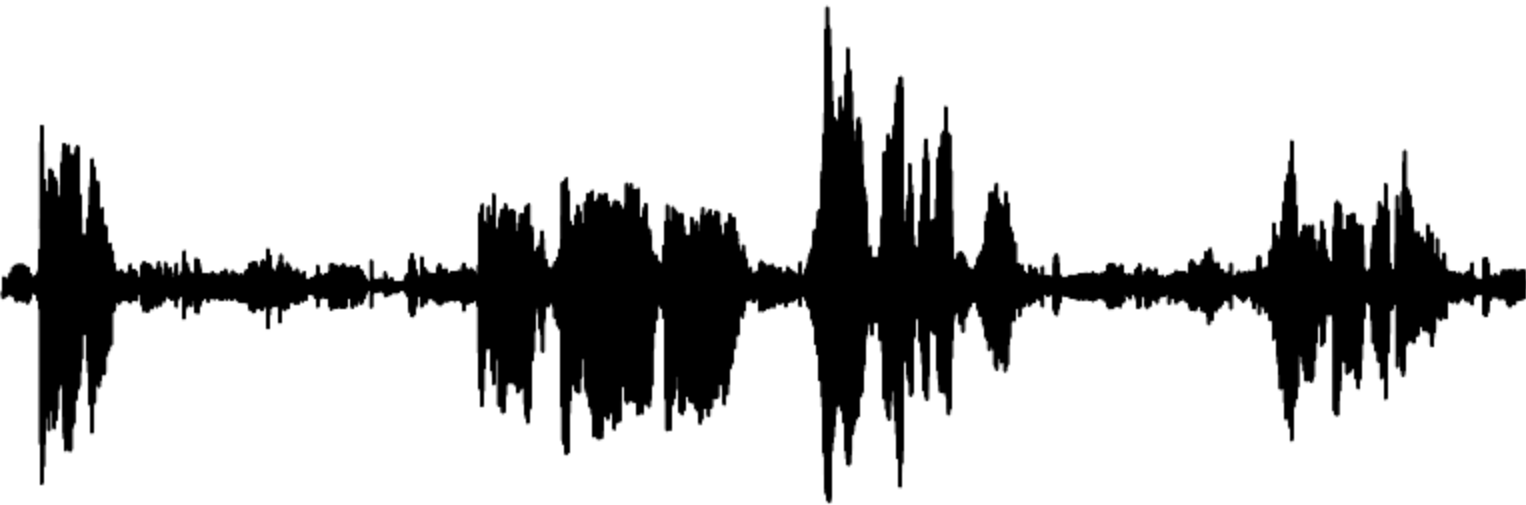} &
        \includegraphics[width=0.18\columnwidth,height=0.5cm]{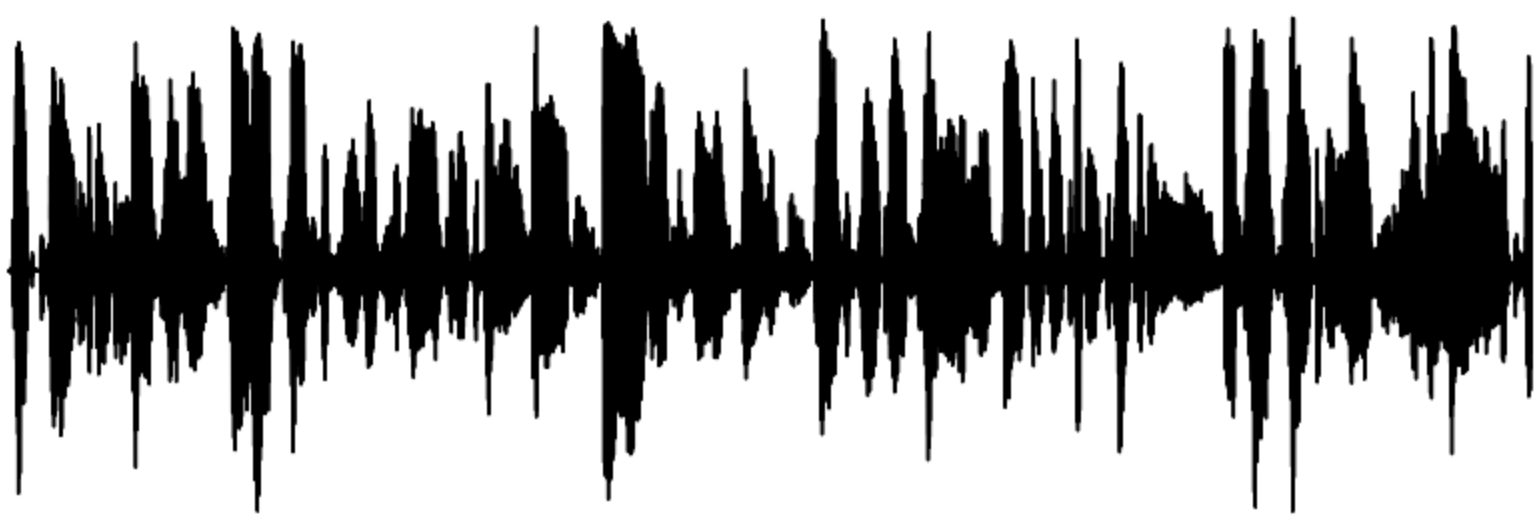} &
        \includegraphics[width=0.18\columnwidth,height=0.5cm]{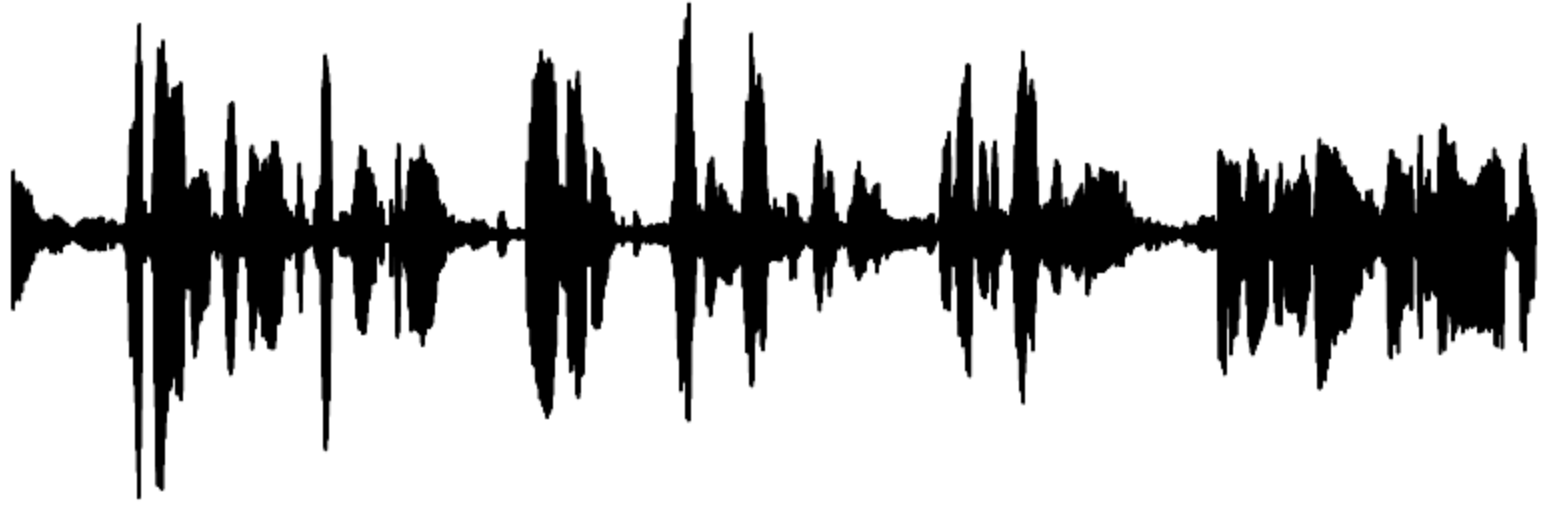} &
        \includegraphics[width=0.18\columnwidth,height=0.5cm]{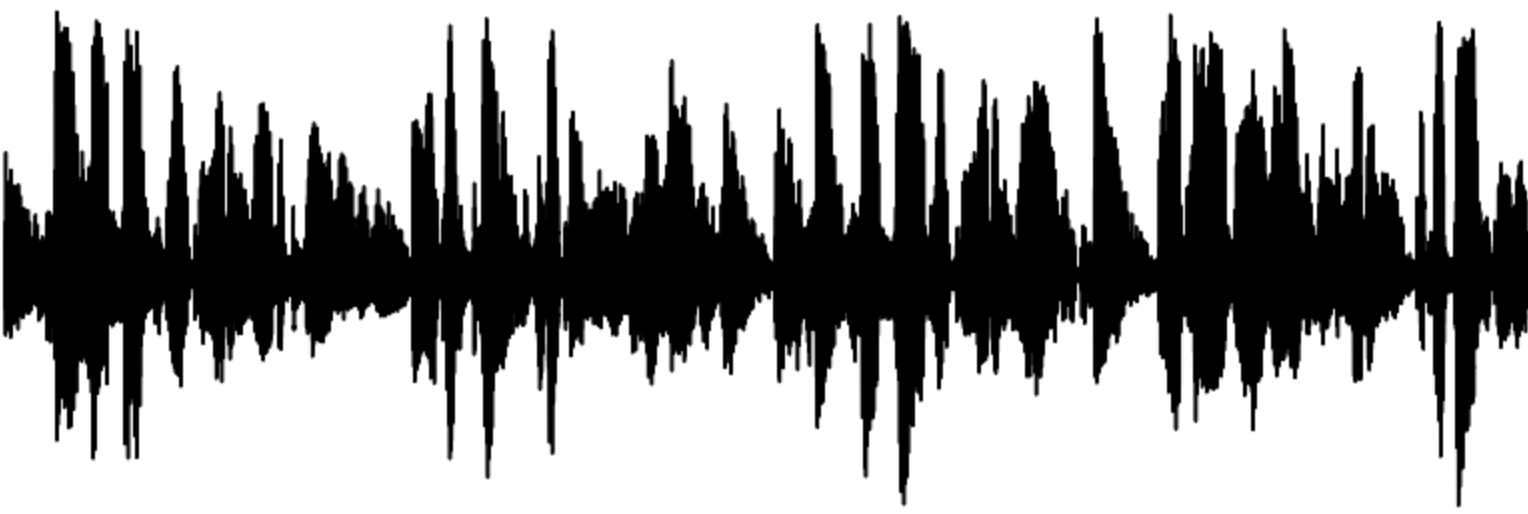} &
        \includegraphics[width=0.18\columnwidth,height=0.5cm]{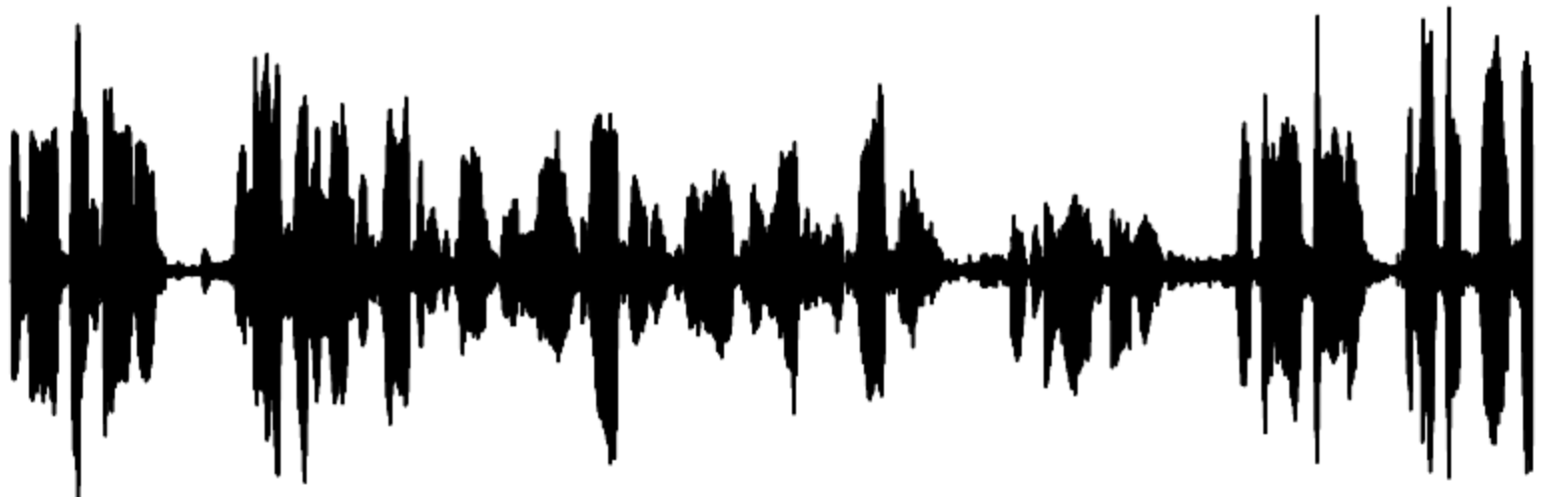} \\

    \end{tabular}
    \caption{%
        Fake samples from the proposed \dbname dataset
       }
    \label{fig:teaser}
    \vspace{-0.8cm}
\end{figure}





To this end, in this paper we contribute a new deepfake dataset, which can be seen as continuation of the work initiated by M\"uller \etal \cite{muller2022does}--the first to highlight the importance of in-the-wild data.
Our dataset, named \dbname, was collected from YouTube (but some of the samples originated from TikTok, Instagram or Facebook) and it is based on fake videos created for both disinformation and entertainment purposes (see Figure~\ref{fig:teaser}).
We use only the audio tracks. The dataset totals 13 hours and comprises samples in eight languages.

We find that the performance of state-of-the-art deepfake detection methods on \dbname is well below any other scientific dataset, or even that on the In-the-Wild dataset~\cite{muller2022does}.
We also note that while there is a proliferation of deepfake detection methods, the most performant ones are in essence similar, relying on strong self-supervised learning (SSL) derived features~\cite{junichi_vocoders,combei_asv2024, martindonas24_interspeech}.
For this reason, instead of over-complicating the model, we take an orthogonal approach and investigate data-centric methods.
Data-centric methods have been successfully employed for a wide range of machine learning problems~\cite{zha2023data,yang2023not,Kumar2024Opportunities},
yet have received only little attention in the deepfake detection community~\cite{azeemi23_interspeech,song2024quality}.
We investigate both data- and algorithm-informed sample selection and pruning methods and
show that they are a remarkably easy and efficient solution for the challenging out-of-domain, real-world samples.

\begin{table*}[ht!]
    \scriptsize
    \newcommand{\na}{{\color{gray}\textsc{n/a}}}
    \newcommand{\ii}[1]{{\scriptsize \color{gray} #1}}
    \newcommand{\lbl}[1]{{\scriptsize \color{gray} #1}}
    \newcommand{\key}[1]{{\footnotesize \texttt{#1}}}
    \newcommand{\dur}[2]{#1{\scriptsize ±#2}}
    \centering

    \caption{%
        An overview of the selected deepfake datasets. 
    }
    \label{tbl:datasets}
    \begin{tabularx}{\textwidth}{lcXcrrrrr}
\toprule
Dataset                          &Year             & Real data                          & Langs.     & Systems        & Real        & Fake        & Duration \\
\lbl{short name}                 &\lbl{release}    &                                    &            &               & \lbl{count} & \lbl{count} & \lbl{seconds} \\
\midrule
ASV19~\cite{wang24_asvspoof}      &2019     &VCTK                                          & en        & 17 TTS/VC        & 7k          & 63k         & \dur{3.1}{1.4} \\
FoR~\cite{for}                   &2019     &Arctic, LJSpeech, VoxForge, YouTube, TED Talks & en         & 6 TTS              & 34k         & 34k         & \dur{3.0}{2.3} \\
ASV21~\cite{asv21}                    &2021     &VCTK                                      & en         & 100 TTS/VC         & 22k         & 589k        & \dur{2.9}{1.2} \\
TIM~\cite{timittts}              &2023     &VidTIMIT                                       & en         & 12 TTS             & 430         & 20k         & \dur{3.1}{1.2} \\
ODSS~\cite{odss}                 &2023     &VCTK, Hi-Fi TTS, HUI-ACG, SLR-ES               & en, es, de & 2 TTS              & 11k         & 19k         & \dur{3.1}{2.0} \\
MLAAD v5~\cite{muller2024mlaad}     &2024     &M-AILABS                                    & many (38)& 82 TTS               & 80k         & 154k        & \dur{8.2}{4.5} \\ 

ASV5~\cite{wang24_asvspoof}     &2024     &MultiLingual LibriSpeech                        &  en       & 13 TTS, 3 VC        & 50k         & 183k        & \dur{9.5}{2.3} \\ 
\hline 
ITW~\cite{muller2022does}        &2022     &YouTube                                        & en         & \na                & 20k         & 12k         & \dur{4.2}{3.3} \\
\dbname                          &2025     &YouTube, Instagram, TikTok, Facebook                     & many (8) & \na          & 3k         & 2k       & \dur{10.0}{0.0}\\

\bottomrule
    \end{tabularx}
\end{table*}

To summarize, our contributions are as follows:
\textbf{(i)} we present a novel, diverse and challenging \textbf{dataset of real-world audio deepfakes}, curated from online platforms;
\textbf{(ii)}~we demonstrate that state-of-the-art \textbf{deepfake detection} systems \textbf{struggle with these real-world samples}, exposing a critical performance gap;
\textbf{(iii)} we show that \textbf{data-centric strategies substantially enhance the detection performance} without the need to adapt the underlying model architecture.

\section{Audio deepfake datasets}

\subsection{Scientific datasets}
\label{sec:datasets}
To address the challenge of detecting real-world deepfakes through data-centric strategies, we start from a wide range of readily available \emph{scientific datasets}. 
We prioritised datasets in languages that allowed our team of fact-checkers and engineers to conduct subjective evaluations. A comprehensive summary of the dataset pool is provided in Table~\ref{tbl:datasets}. We note here only the differences with respect to their complete releases.
From \textbf{Fake or Real (FoR)}~\cite{for} we utilize the \texttt{for-norm} partition.
For \textbf{ASV5}, due to its late full release in mid-December 2024, we limited our selection to the training and development subsets alone.
From \textbf{ASV21}, not wanting to bias the results based on its large size, we randomly selected approximately 60k samples, ensuring the same distribution as the full dataset.
To augment the \textbf{MLAAD v5} dataset~\cite{muller2024mlaad} with genuine samples, we selected 80k real samples from the M-AILABS dataset~\cite{mailabs}.

\subsection{Real-world datasets}

As baseline real-world samples we start from the \textbf{In-the-Wild (ITW)}~\cite{muller2022does} dataset. It includes a wide variety of content sourced from multiple platforms, environments, and manipulation techniques. The data generation methods of ITW are unknown. 

A second real-world dataset is our own contribution and we refer to it as \textbf{\dbname}. 
\dbname dataset includes 196 fake and 192 real videos in 8 languages, with a total duration of around 13~hours. The videos were collected from YouTube, but some of these samples were originally released on other platforms (e.g. Facebook, Instagram or TikTok) over the last two years.
The fake samples are diverse in terms of the target audience: political disinformation and financial scams; or harmless entertainment using public and political figures.
\dan{The samples were identified as fake either based on their metadata or by journalist fact-checkers.}
\dan{The real samples were downloaded from YouTube and maintain similar content to the fakes.}
To control for length, 
we segment the original audio into \dan{ten-second} chunks, resulting in 2005 fake and 2793 real segmented samples.
The complete list of audio sources is available in our code repository.

\dan{
By manually inspecting the samples in these two datasets,
we find they lack major artefacts, such as mispronunciations and generation errors.
This suggests that the samples have been curated to increase credibility and alignment with a specific goal.
}

\section{Baseline deepfake detection system}
\label{sec:baseline}

\begin{table}[t]
\centering
\setlength{\tabcolsep}{2pt}
\newcommand{\na}{{\color{gray}\textsc{n/a}}}
\newcommand{\ii}[1]{{\scriptsize \color{gray} #1}}
\scriptsize
\caption{
    EER $\downarrow$ evaluation for our baseline SSL-based deepfake detection model.
    ASV19 train+dev subsets are used for training, and rows represent the test data.
    SotA column reports best results from previous works.
    (L) refers to the last hidden state, and (B:*) refers to the best SSL layer indicating their number.
    R is the RawBoost augmentation, C is the codec augmentation and RB+C is the RawBoost plus codec augmentation. Best results are highlighted.
    Standard deviations are reported only for the columns which involve random selection of data.
}
\label{tbl:baseline}
\begin{tabular}{l|l|r|rr||r|r|r}
\toprule

&\textbf{Dataset}& \multicolumn{1}{c|}{SotA} & \multicolumn{2}{c||}{\texttt{xls-r-2b}}&                                   \texttt{+RB} & \texttt{+C} & \texttt{+RB+C}  \\ 

&                &                                                               & (L) & (B:9)                           &  &  &                                                    \\ \midrule
\multirow{7}{*}{\rotatebox{90}{\ii{scientific}}}&ASV19-eval& 0.1~\cite{junichi_vocoders} & 0.6 & 0.1  &                    \cellcolor{blue!25} \textbf{0.07±0.01}      &   0.1±0.02                       & 0.08±0.0\\
&FoR             & 6.9~\cite{pascu_is24}                                         & 6.6 & 6.6&                               6.3±0.3        &  6.4±0.3                       & \cellcolor{blue!25} \textbf{6.0±0.2}\\
&ASV21           & \cellcolor{blue!25}\textbf{2.1}~\cite{truong24b_interspeech}   & 3.3 & 2.3&                               2.3±0.1         & 2.4±0.06                        & 2.4±0.06\\
&TIM             & 11.5~\cite{pascu_is24}                                        & 15.2& \cellcolor{blue!25}\textbf{5.6}&                               9.9±1.5         &10.9±2.5 & 12.7±1.5\\
&ODSS            &\cellcolor{blue!25}\textbf{16.0}~\cite{pascu_is24}              & 17.0& 16.2&                              17.2±1.2   & 16.7±0.4                            & 17.6±0.9\\
&MLAAD v5        &\na                                                            & 17.0& 12.8&   12.6±0.2    & \cellcolor{blue!25}\textbf{12.4±0.4}                             & 12.9±0.2\\
&ASV5            & \na                                                           & 1.7 & 0.9   &                            0.9±0.06    & 0.9±0.01                             & \cellcolor{blue!25}\textbf{0.9±0.01}\\ \midrule
\multirow{2}{*}{\rotatebox{90}{\ii{real}}} &ITW&\cellcolor{blue!25}\textbf{3.1}~\cite{martindonas24_interspeech}& 7.3 & 3.4& 3.5±0.1     & 3.3±0.2                              & 3.4±0.1 \\
&\dbname         & \na                                                            & 34.2& \cellcolor{blue!25}\textbf{27.4}&  27.7±0.4   & 27.5±0.4                             & 28.0±0.9\\ \midrule

 &\textbf{Mean scientific} & \na                                                 & 8.8&  \cellcolor{blue!25}\textbf{6.4}  &                             7.0±0.5& 7.1±0.5 & 7.5±0.2 \\ 
                                            &\textbf{Mean real}   & \na  & 20.8& \cellcolor{blue!25}\textbf{15.4 }&          15.3±0.3 & 15.3±0.3 & 15.7±0.5\\ 
                                            &\textbf{Mean all}   & \na   & 11.5& \cellcolor{blue!25}\textbf{8.4} &           8.9±0.4 & 9.0±0.4 & 9.3±0.2 \\ \bottomrule
\end{tabular}
\vspace{-.1cm}
\end{table}

\dan{
We benchmark the detection performance on the newly introduced AI4T dataset.
To this end, we build a strong baseline model based on the most promising ideas in deepfake detection.
The main ingredient is the use of self-supervised learnt (SSL) audio representations,
which have shown strong performance on this task \cite{junichi_vocoders,martindonas24_interspeech,pascu_is24}.
We then improve this by using intermediary SSL representations and data augmentation.
}

\dan{
\textbf{Initial model.}
Given an audio file, we extract features from the last hidden state of the frozen wav2vec2 XLS-R 2B model.%
\footnote{\url{https://huggingface.co/facebook/wav2vec2-xls-r-2b}}
These features are then average pooled across time yielding a 1,920-dimensional representation.
The back-end is a logistic regression classifier trained with weak regularisation ($C=10^{6}$).
Following prior work \cite{pascu_is24}, we train on the ASV19 training and development sets and report Equal Error Rate (EER).
The results in Table \ref{tbl:baseline} show good performance on most of the datasets with the exception of AI4T where the EER is 34.2\%.
}

\dan{
\textbf{Intermediary representations.}
Instead of relying on the last-layer representations,
it was showed that intermediary representations provide better features for deepfake detection \cite{martindonas24_interspeech,zhang2024audio}.
We exhaustively explore all layers and find that the best performance on scientific datasets is obtained at layer~9 (out of 48). 
In Table \ref{tbl:baseline} (column B:9), we observe significant performance improvements over the last layer (column L) for most of the datasets, including ITW and \dbname.
However, the performance gap between the two real-world datasets remains.
}

\textbf{Data augmentation.}
To further improve the performance of our baseline,
we perform data augmentation
\cite{tomilov2021stc, chen2021pindrop, wang2023spoofed,Tak2021RawboostAR, tak2022automatic, martindonas24_interspeech, zhang2024audio, truong24b_interspeech}:
we augment the training samples using noise (RawBoost \cite{Tak2021RawboostAR}) and audio codecs (OPUS and AAC).
We partition the dataset into the number of augmentations and apply a single augmentation to each partition.
The results in Table \ref{tbl:baseline} (last three columns) show that data augmentation improves the detection for the scientific datasets,
but, somewhat surprisingly, no performance gain is observed on the real-world datasets (ITW and \dbname).

These results show that state-of-the-art deepfake detection suffers a significant performance drop when moving from scientific data to the \dbname real-world dataset.
Despite this gap, we argue that scientific datasets contain useful information.
Next, we search for this information using data-centric approaches,
which systematically refine the selection of training samples.

\begin{table}[t!]
\centering
\newcommand{\ii}[1]{{\scriptsize \color{gray} #1}}
\newcommand{\lbl}[1]{{\scriptsize \color{gray} #1}}
\setlength{\tabcolsep}{.7pt}
\scriptsize

\caption{Dataset mixing EER $\downarrow$ results. The table lists the best results for each subset of $N$ datasets along with the baseline.
}
\label{tbl:mixing}
\begin{tabular}{cccccccc|rr|r}
\toprule
\ii{\# datasets} &\rotatebox{0}{ASV19} & \rotatebox{0}{FoR} & \rotatebox{0}{ASV21} & \rotatebox{0}{TIMIT} & \rotatebox{0}{ODSS} & \rotatebox{0}{MLAAD} & \rotatebox{0}{ASV5} & \textbf{ITW} & \textbf{AI4T} & \textbf{Mean} \\
\midrule
\ii{baseline}&\checkmark &        &        &        &        &           &        & 3.41	   &27.43 &	15.42  \\ \midrule
\ii{1}&           &        &        &        &\checkmark&       &        & 4.54	&17.24 & 10.89 \\
\ii{2}&\checkmark&         &        &        & \checkmark&      &        & 3.77 &14.28 & 9.03    \\
\ii{3}&           &\checkmark&       &        & \checkmark&\checkmark&        & 3.07 &	\cellcolor{blue!25}\textbf{13.32} &	\cellcolor{blue!25}\textbf{8.20}         \\    
\ii{4}& & \checkmark&      &\checkmark&        &\checkmark&\checkmark& \cellcolor{blue!25} \textbf{2.55}	&14.14	& 8.35      \\
\ii{5}&            &\checkmark&        &\checkmark&\checkmark&\checkmark&\checkmark& 2.64&	13.88 &	8.26        \\
\ii{6}&\checkmark&        &\checkmark&\checkmark&\checkmark& \checkmark&\checkmark&       4.96 &	18.85&	11.91   \\
\ii{7}&\checkmark&\checkmark&\checkmark&\checkmark&\checkmark&\checkmark&\checkmark& 2.90&	29.03&	15.97  \\

\bottomrule
\end{tabular}

\vspace{-.5cm}
\end{table}

\section{Data-centric strategies}

Data-centric strategies aim to simplify the computational complexity of a machine learning task through the optimization and enhancement of the data, rather than by modifying the model architecture or algorithm~\cite{press2021andrew,zha2023data,Kumar2024Opportunities}. The idea is that high-quality, well-processed data is often more impactful than complex models in achieving better generalization and accuracy.
We do this in the context of audio deepfake detection by exploiting the large number of scientific datasets through two main approaches: at dataset- and at individual sample-level.

\subsection{Scientific dataset selection}

In Section~\ref{sec:baseline} we showed that when it comes to more recent deepfake samples (i.e., the \dbname dataset), the contents of ASV19 is unable to provide a good generalisation performance for the underlying model.
However, as seen in Section~\ref{sec:datasets}, there are many other datasets that could be used for training.
In this section we aim to find the combination that yields the best results.
We consider the seven scientific datasets and evaluate all $128$~($2^7$) combinations.
For each dataset we use all of its samples, including real samples and those in the test splits.
We use layer 9 features and no data augmentation.

Table~\ref{tbl:mixing} presents 
the best results for each combination of $N$ datasets (single dataset, pairs of datasets, etc.);
full results are available in the accompanying code repository.
We notice that for ITW, the ASV19-based baseline is already at a good classification performance (3.41\% EER), and that the best combination of datasets yields a 1\% absolute EER improvement down to 2.55\% (line 4). 
However, for the \dbname dataset almost all dataset mixes yield large improvements, with the best result halving the initial EER to 13.32\% (line 3 in Table~\ref{tbl:mixing} composed of FoR, ODSS and MLAAD datasets). 

\begin{figure}
    \centering
    \includegraphics[width=\linewidth]{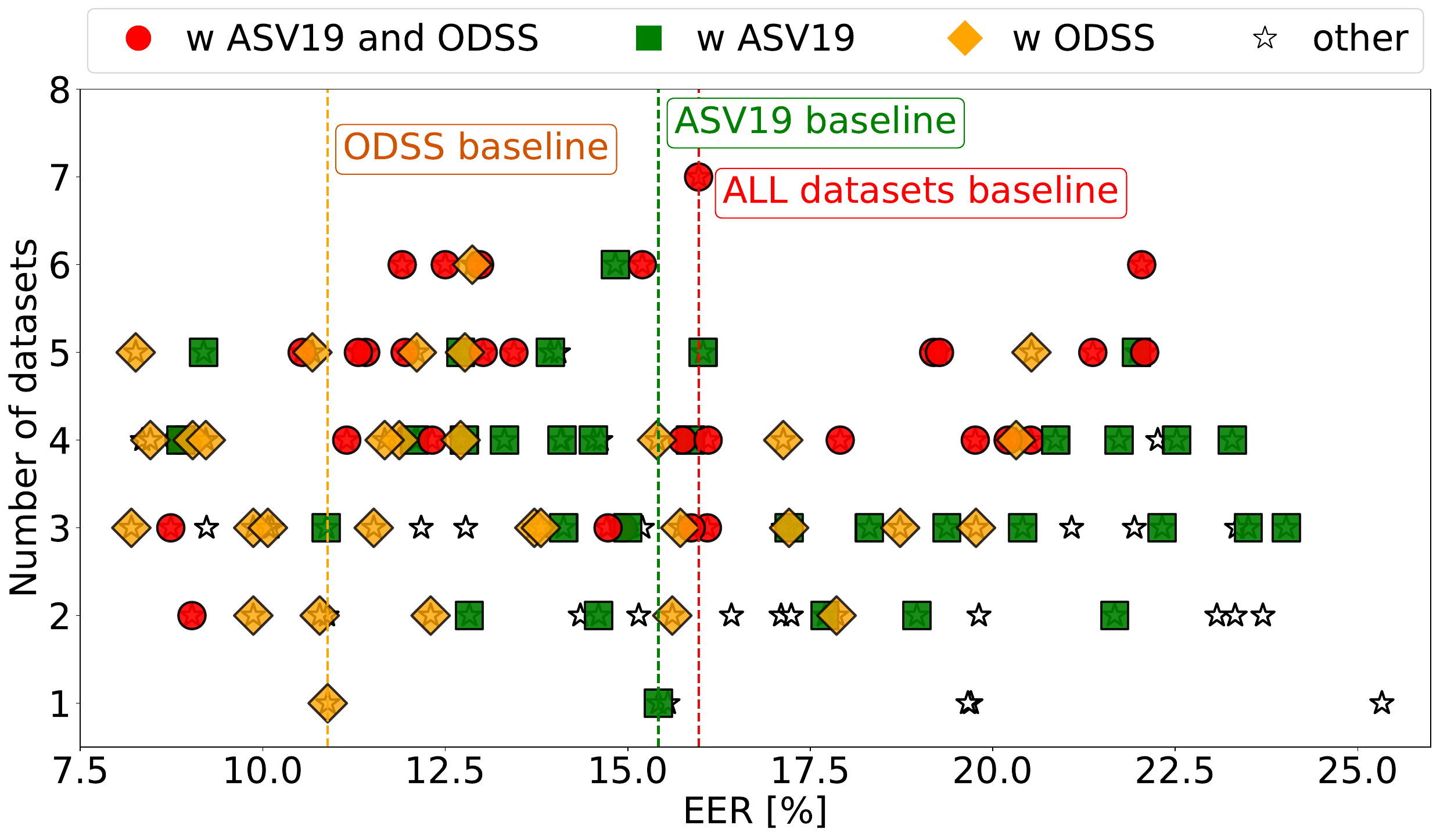}
     \caption{Average EER $\downarrow$ performance for the $2^7$ dataset combinations. We use different markers to denote the combinations which include: ASV19 (\(\square\)), ODSS (\(\diamondsuit\)), or both ASV19 and ODSS ($O$).
    }
    \label{fig:db-mixing-eer-vs-num-datasets}
    \vspace{-.5cm}
\end{figure}

To further analyse the results, we plot in Figure~\ref{fig:db-mixing-eer-vs-num-datasets} the performance of all 128 combinations with respect to the number of datasets used for training 
We observe that the error does not decrease with the number of datasets and that there is a large variance at each level.
These observations suggest that the characteristics of the training datasets are critical for good performance,
and some of the datasets might even hurt performance.
As an example, using all datasets yields an average EER across the two real-world datasets of 15.97\% which is very close to using the popular ASV19 dataset (15.42\% EER).
If we focus only on the combinations that include ASV19 (green squares and red circles),
we see that about about half of the results are better and half are worse than the single-dataset baseline. 
The best single performing dataset is ODSS:
ODSS yields an EER of 10.89\% on its own, and this performance can be improved to 8.20\% when including also MLAAD and FoR.
However, most of the combinations that include ODSS (orange diamonds and red circles) have worse performance than the single dataset.




\begin{figure}[t!]
    \centering
    \includegraphics[trim=0 0 0 0, clip, width=\linewidth]{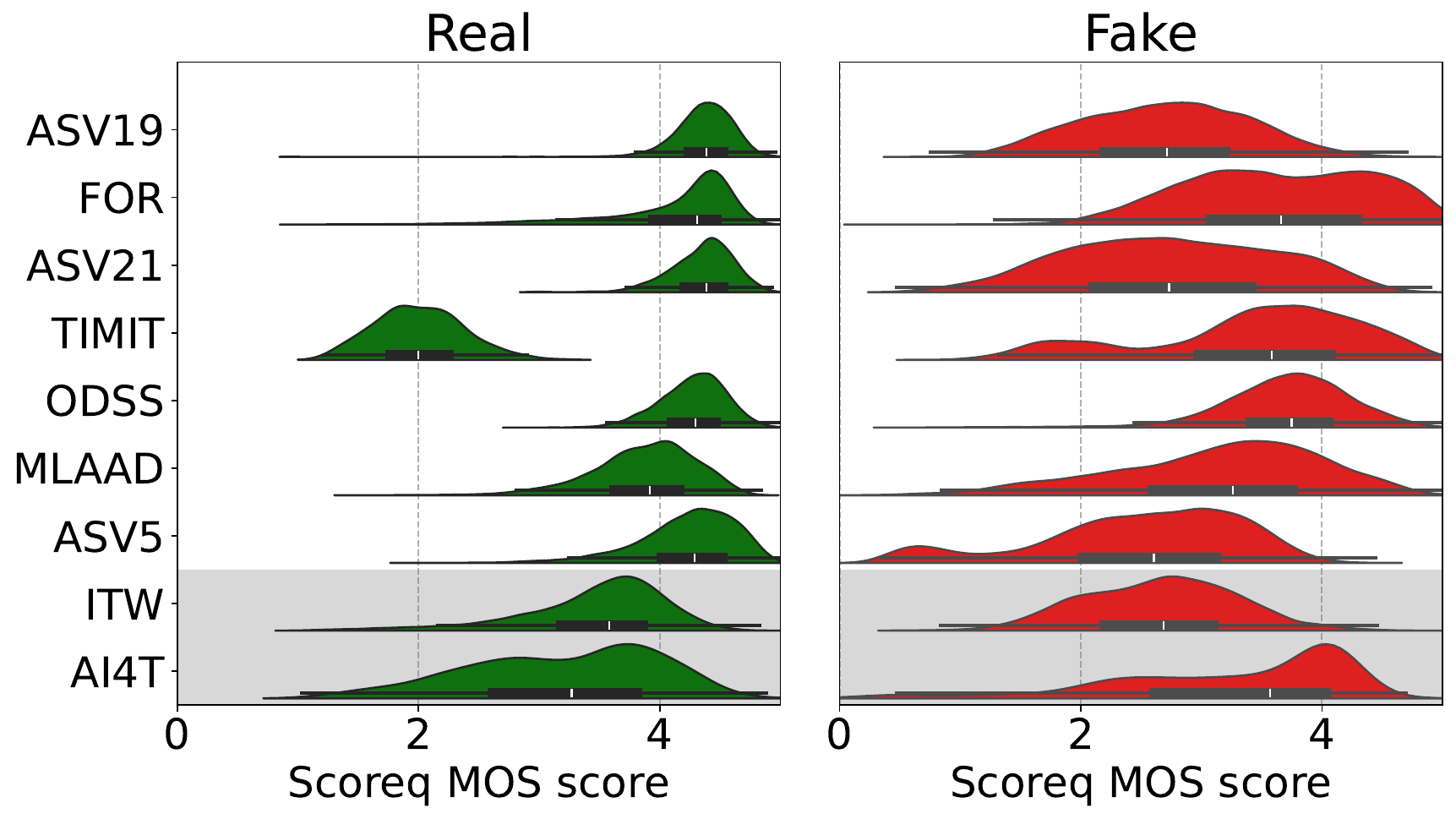}
    \caption{%
       Distribution plots of the automatic Scoreq MOS scores $\uparrow$ over the real and fake samples from all datasets.
    }
    \label{fig:scores}
    \vspace{-.5cm}
\end{figure}

Why do some datasets help, while some other hurt performance?
An important aspect is the chronology of these datasets (see Table~\ref{tbl:datasets}):
ITW is closer in time to ASV19, ASV21 and FoR, while \dbname is closer to ODSS, MLAAD and ASV5. 
The date of release may reflect the dataset quality.
We verify this hypothesis by automatically estimating the speech quality with Scoreq~\cite{ragano2024scoreq}.
The distributions in Figure~\ref{fig:scores} indicate that ITW correlates well with the ASV19 and ASV21 data for the fake samples, and with MLAAD (M-AILABS, to be precise) for the real samples. 
However, \dbname has a much wider score distribution for the real samples 
than any of the scientific datasets, and a rather high quality for the fake samples
--lightly correlated to ODSS and FoR. 
FoR also contains real data collected from online platforms and fake samples from high-quality commercial speech synthesisers, which may partly explain the increased performance when using it for training. 


\begin{figure}[t!]
  \centering
  \scriptsize
  \includegraphics[width=\columnwidth,trim={0pt 0pt 0pt 0pt},clip]{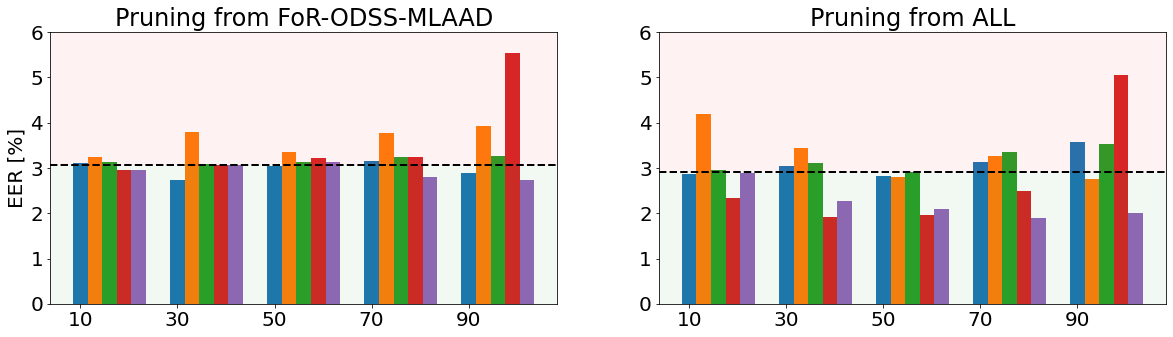}\\
  (a) ITW
  \includegraphics[width=\columnwidth,trim={0pt 0pt 0pt 0pt},clip]{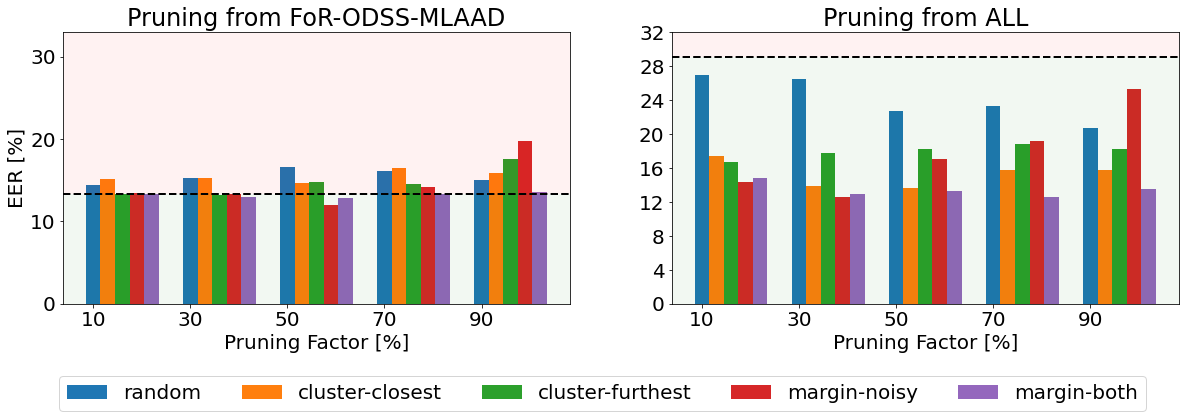}\\
  (b) \dbname
  \caption{%
    Pruning results on the two real-world datasets using five methods,
    when starting from the best dataset combination (left) or all datasets (right). 
    The horizontal line indicates the performance with no pruning.
    The pruning factor is the percent of discarded samples. 
    For random pruning we report average results over three random seeds.
  }
  \label{fig:pruning}
  \vspace{-.5cm}
\end{figure}

  

\subsection{Sample pruning}

Although the deepfake datasets were released as cohesive units, not all their samples may be equally relevant for the generalisation objective. 
As a result, similar to \cite{azeemi2022representative,azeemi23_interspeech,10389626}, we explore three data pruning strategies:
(i) random pruning; (ii) data-informed pruning; and (iii) algorithm-informed pruning.
Random pruning refers to randomly discarding samples from the training data, which was shown to exhibit good generalisation when no sample scoring can 
used~\cite{pruning,drop}. 
For data-informed pruning, we rank samples by the distance to their mean, and select either the closest (\texttt{cluster-closest}) or the furthest (\texttt{cluster-furthest}) samples.
The samples are represented by their average pooled self-supervised embeddings and compared using the Euclidean distance.
We apply this strategy independently on both real and fake samples from each of $N$ datasets,
and obtain $2N$ sets of samples, which we ensemble in the final pruned dataset.
For algorithm-informed pruning, we use the logistic regression's margins over the samples.
We remove the closest (\texttt{margin-noisy}) or the closest and furthest (\texttt{margin-both}) points with respect to the decision hyperplane, irrespective of the dataset that they belong to.

Figure~\ref{fig:pruning} shows a subset of the results of the sample pruning methods (complete results are available in the code repository).
We use either the best combination of datasets (i.e. \texttt{For+ODSS+MLAAD}) or the complete set of available samples (\texttt{ALL} data) as the starting pool for the pruning strategies, and report the EER over the real-world datasets.  
First, we observe that performance is relatively stable with the amount of discarded data.
This means that many of the samples carry redundant information.
For the data pruned from the FoR+ODSS+MLAAD combination (left side of the figure), the average results obtained after pruning are similar to the baseline.
The only pruning method that performs marginally better is \texttt{margin-both}. 
However, obtaining a similar performance when discarding 90\% of the data is in itself a very important observation, enabling a more efficient model selection.
When pruning the complete set of samples (right side of the figure), margin-based selections work best for both ITW and \dbname.
For \dbname, almost all strategies halve the baseline EER (bottom right plot).
This may be because the baseline for \dbname is rather poor, at 29.03\% EER,
and any noisy and outlier samples' reduction are relevant to the logistic regressor.

The best results obtained for ITW starting from the two collections of samples are: 2.74\% EER for the For-ODSS-MLAAD subset using \texttt{margin-both} at 90\% pruning factor; and 1.70\% EER for the ALL subset using \texttt{margin-both} strategy at 80\% pruning factor. The corresponding \dbname results are 13.53\% and 12.43\% EER, respectively. The 1.70\% EER obtained for ITW represents a 
55\% relative increase in performance over the state-of-the-art results,~i.e.~3.1\% reported by Martin-Donas et al.~\cite{martindonas24_interspeech}.

\subsection{Post-pruning data augmentation}

Having selected the best combination of datasets and their most representative samples, we go back to the initial data augmentation strategy in hope of improving the results even further. 
For all entries reported in Table~\ref{tbl:augm} we adopt the same data augmentation strategy as in Section~\ref{sec:baseline}. The \emph{No augmentation} line reports the best results obtained in the previous sections. Given the randomness of the sample augmentation, we report mean and standard deviations over 3 random seeds.  
As opposed to Table~\ref{tbl:baseline} where no gain was observed for the real-world datasets, the results over the pruned datasets show a 2\% absolute increase in the performance for the \dbname dataset. There is a relatively limited or no improvement over ITW using this data augmentation strategy, which may be partly caused by the irreducible error within the dataset.
These results 
show that data augmentation can help with generalisation, but only if the underlying data is of sufficient quality.

\begin{table}[t!]
\centering
\footnotesize
\newcommand{\ii}[1]{{\scriptsize \color{gray} #1}}
\setlength{\tabcolsep}{2pt}
\caption{Data augmentation EER $\downarrow$ performance after dataset selection and sample pruning. The augmentation is performed over the samples selected from either the best combination of datasets (For+ODSS+MLAAD), or from the entire set of scientific samples (ALL). 
}
\label{tbl:augm}
\vspace{-.1cm}
\begin{tabular}{l|rr|rr}

\toprule
       & \multicolumn{2}{c|}{\textbf{FoR+ODSS+MLAAD}} & \multicolumn{2}{c}{\textbf{ALL}} \\ \midrule
       \ii{Augm. method}                        & \textbf{ITW} & \textbf{\dbname} & \textbf{ITW }& \textbf{\dbname} \\ \midrule
     No augmentation           &  2.7   &   13.5    &  \cellcolor{blue!25}\textbf{1.70}    &   12.4   \\  \midrule
     +RawBoost                &  2.5±0.1    &   11.5±0.3    &  2.2±0.1                               &  11.7±0.3   \\  
     +Codecs                  &  2.7±0.3    &  14.0±0.1     &  2.0±0.1                              &    12.0±0.7  \\
     +RawBoost+Codecs         & \cellcolor{blue!25}\textbf{2.4±0.4} &  \cellcolor{blue!25}\textbf{11.3±0.3}  & 1.9±0.2    &  \cellcolor{blue!25}\textbf{10.2±0.2}    \\\midrule

\end{tabular}
\end{table}

\vspace{-.2cm}
\section{Conclusions}

Our results have shown that scientific datasets, while seemingly disjoint from real-world deepfake samples, still contain essential information that can greatly impact model performance.
Using a data-centric methodology, we were able to achieve a 55\% performance improvement on the ITW dataset (at 1.70\% EER) and a 63\% performance improvement on the newly proposed and challenging \dbname dataset (at 10.2\% EER). 
While we note that different detection methods may require different selection strategies, we argue that it is essential to prioritize data analysis and data-centric approaches before expanding the capacity of the model (which could potentially obscure its explainability and deployment feasibility). 

As future work, we 
will investigate what makes a relevant sample in the context of deepfake detection.
Why are some samples more informative than others? 
How can we detect the common artefacts of the deepfakes, in order to help us trace them better?

{\textbf{Acknowledgement.} This work was co-funded by EU Horizon project AI4TRUST (No. 101070190), and by the Romanian Ministry of Research, Innovation and Digitization project
DLT-AI SECSPP (id: PN-IV-P6-6.3-SOL-2024-2-0312), and by the Free State of Bavaria under the DSgenAI project (Grant Nr.: RMF-SG20-3410-2-18-4).


\bibliographystyle{IEEEtran}
\bibliography{main}

\end{document}